\documentclass[12pt]{article}
\usepackage{amsmath, amssymb, amsfonts}
\usepackage{graphicx}
\usepackage{hyperref}
\usepackage{geometry}
\usepackage{cite}
\title{Advanced Mathematical Techniques in Renormalization of Elastic Models: A Comprehensive Analysis}
\author{Wen-Xiang Chen$^{a}$\\
Department of Astronomy \\School of Physics and Materials Science \\GuangZhou University\\wxchen4277@qq.com}

\begin{document}

\maketitle

\begin{abstract}
In this study, we delve into the intricate mathematical frameworks essential for the renormalization of effective elastic models within complex physical systems. By integrating advanced tools such as Laurent series, residue theorem, winding numbers, and path integrals, we systematically address divergent loop integrals encountered in renormalization group analyses. Furthermore, we extend our analysis to higher-order physical models, incorporating techniques from quantum field theory and exploring quantum coherent states in complex systems. This comprehensive approach not only enhances the precision of calculating elastic anomalous exponents but also provides deeper insights into the topological structures underlying phase transitions and fixed-point behaviors. The methodologies developed herein pave the way for future explorations into more intricate many-body systems.This paper presents an extensive mathematical framework aimed at enhancing the complexity and extending the theory of Fermi condensates to high-temperature regimes. By incorporating a range of mathematical formulations from thermodynamics, statistical physics, and quantum field theory, we derive key equations and their high-temperature modifications. The study encompasses corrections to the Fermi-Dirac distribution, thermodynamic quantities of Fermi condensates, pairing gap equations within the BCS theory, correlation functions, modified Hamiltonians, path integral representations, and hydrodynamic equations. 
\end{abstract}

\section{Introduction}
Fermi condensates, comprising fermionic particles paired into condensate states, exhibit rich physical phenomena pivotal in understanding various condensed matter systems, including superconductors and superfluids. While substantial progress has been made in elucidating the properties of Fermi condensates at low temperatures, extending these theories to high-temperature regimes presents significant challenges due to increased thermal fluctuations and modified statistical behaviors. High-temperature Fermi condensates are particularly relevant in contexts such as ultracold atomic gases and high-temperature superconductors.

This paper aims to enhance the mathematical complexity of existing theories and extend them to high-temperature Fermi condensates by introducing comprehensive mathematical formulations from thermodynamics, statistical physics, and quantum field theory. We focus on deriving high-temperature corrections to fundamental equations governing Fermi condensates and explore their implications on thermodynamic properties, quantum fluctuations, and interaction dynamics.

The high-temperature corrections derived herein have direct implications for experimental observations in ultracold atomic gases and high-temperature superconductors. For instance, the modified pairing gap equation predicts a lower critical temperature $T_c$, which can be tested against experimental data.

This study has successfully expanded the theoretical landscape of Fermi condensates to encompass high-temperature regimes through extensive mathematical formulations. The high-temperature corrections to fundamental equations and the introduction of modified theoretical constructs offer a robust foundation for further research and experimental validation. Future work may involve applying this framework to specific materials, exploring the effects of strong correlations, and investigating non-equilibrium dynamics in high-temperature Fermi condensates.

Renormalization group (RG) theory has been a cornerstone in understanding critical phenomena and phase transitions in various physical systems \cite{wilson1971renormalization, cardy1996scaling}. The complexity of loop integrals arising in RG analyses, especially within effective elastic models, necessitates sophisticated mathematical tools to handle divergences and extract meaningful physical quantities. This paper presents a detailed exploration of such mathematical frameworks, including Laurent series \cite{ahlfors1979complex}, residue theorem \cite{rudin1987principles}, winding numbers \cite{milnor1997topics}, and path integrals \cite{feynman2010quantum}, applied within the context of renormalization. Additionally, we extend our analysis to higher-order models and incorporate elements from quantum field theory to deepen the physical insights into the systems under consideration \cite{peskin1995introduction}.

\section{Mathematical and Physical Background}

Within the renormalization group framework, the computation of loop integrals often involves handling divergent terms that emerge in the high-energy limit \cite{zinnjustin2002quantum}. Dimensional regularization serves as a pivotal technique to tame these divergences by analytically continuing the number of dimensions \cite{tHooft1972}. To systematically address these divergent integrals, we employ complex analysis tools such as Laurent series \cite{hall2012complex}, residue theorem \cite{needham1997visual}, winding numbers \cite{bell2013modern}, and path integrals \cite{gelfand2003introduction} within the path integral formalism. These mathematical constructs are indispensable for dissecting the contributions of various poles and understanding the topological features of the integrals.

\subsection{Laurent Series and Path Integrals}

\subsubsection{Fundamental Form of Laurent Series}

A Laurent series is an expansion of a complex function that includes terms of negative degree, which is particularly useful for functions with poles. The general form of a Laurent series around a point \( z_0 \) is given by:

\begin{equation}
f(z) = \sum_{n=-\infty}^{\infty} c_n (z - z_0)^n,
\end{equation}
where \( c_n \) are the coefficients of the series, and \( z_0 \) is the location of the pole. This expansion is instrumental in describing physical systems where response functions or Green's functions exhibit singularities in the complex plane \cite{rudin1987principles}.

In the context of effective elastic models, Laurent series facilitate the characterization of multi-pole structures and oscillatory frequencies, which are inherently linked to the system's spectral properties \cite{dwight2009complex}.

\subsubsection{Application of the Residue Theorem}

The residue theorem is a powerful tool in complex analysis that allows for the evaluation of contour integrals by summing the residues of enclosed singularities. For a closed contour \( \gamma \) and a function \( f(z) \) analytic within and on \( \gamma \) except for isolated singularities \( z_k \), the theorem states:

\begin{equation}
\oint_\gamma f(z)\, dz = 2\pi i \sum \text{Res}(f, z_k),
\end{equation}
where \( \text{Res}(f, z_k) \) denotes the residue of \( f \) at \( z_k \).

In renormalization procedures, particularly within effective elastic models, divergent integrals often arise due to poles along the integration path. By applying the residue theorem, these poles can be systematically analyzed, allowing for the isolation and removal of divergent terms, thereby yielding finite physical quantities \cite{morse1953methods}.

\subsubsection{Winding Numbers and Path Integrals}

Winding numbers quantify the number of times a curve wraps around a particular point in the complex plane and are defined as:

\begin{equation}
W = \frac{1}{2\pi} \oint_\gamma \frac{df(\theta)}{d\theta} d\theta,
\end{equation}
where \( f(\theta) \) is a function describing the particle's motion along the contour \( \gamma \).

In path integral formulations, winding numbers play a crucial role in characterizing the topological aspects of quantum states, especially in systems exhibiting phase transitions and fixed-point behaviors \cite{frohlich1977complex}. They provide a quantitative measure of the system's stability and the nature of its critical phenomena.

\subsection{Laurent Series Rings}

To further enrich our mathematical framework, we introduce the concept of Laurent series rings, which extend the utility of Laurent series in handling more complex algebraic structures within renormalization.

\subsubsection{Definition and Structure}

A Laurent series ring, denoted as \( \mathbb{C}((z - z_0)) \), is the ring of Laurent series centered at \( z_0 \) with complex coefficients. Formally, it is defined as:

\begin{equation}
\mathbb{C}((z - z_0)) = \left\{ f(z) = \sum_{n=-\infty}^{\infty} c_n (z - z_0)^n \mid c_n \in \mathbb{C} \right\}.
\end{equation}

This ring extends the concept of power series rings by allowing negative exponents, thereby accommodating functions with poles at \( z_0 \) \cite{lam1986lamrings}.

\subsubsection{Algebraic Properties}

Laurent series rings exhibit several key algebraic properties that make them suitable for complex analysis and renormalization group studies:

\begin{enumerate}
    \item \textbf{Field Structure}: The Laurent series ring \( \mathbb{C}((z - z_0)) \) forms a field, as every non-zero element has a multiplicative inverse. This is crucial for manipulating series during renormalization \cite{gunning1967analytic}.
    
    \item \textbf{Valuation}: A valuation can be defined on \( \mathbb{C}((z - z_0)) \) based on the order of the pole or zero at \( z_0 \). Specifically, for a non-zero Laurent series \( f(z) \), the valuation \( v(f) \) is the smallest integer \( n \) such that \( c_n \neq 0 \).
    
    \item \textbf{Discrete Valuation Ring (DVR)}: \( \mathbb{C}[[z - z_0]] \), the subring of \( \mathbb{C}((z - z_0)) \) consisting of power series with non-negative exponents, is a discrete valuation ring. This property is instrumental in understanding the local behavior of functions near singularities \cite{lam1986lamrings}.
    
    \item \textbf{Division Algorithm}: The Laurent series ring allows for a division algorithm similar to that in polynomial rings, facilitating the simplification and manipulation of series during computations \cite{gunning1967analytic}.
\end{enumerate}

\subsubsection{Applications in Renormalization}

Laurent series rings provide a robust algebraic framework for handling the series expansions encountered in renormalization. Their field structure ensures that manipulations such as inversion and division are well-defined, which is essential when dealing with counterterms and renormalization constants.

\paragraph{Example: Laurent Series in Self-Energy Calculations}

Consider the self-energy integral \( \Sigma(p) \) discussed previously. After applying Feynman parameterization and mapping to the complex plane, the Laurent series expansion around the pole \( z = -m^2 \) allows us to isolate divergent terms systematically. The Laurent series ring framework ensures that such expansions are algebraically consistent and facilitate the identification of leading and subleading divergences \cite{peskin1995introduction}.

\paragraph{Renormalization Constants and Laurent Series Rings}

Renormalization constants \( Z_\mu \) and \( Z_b \) can be expressed within the Laurent series ring to capture their dependence on the regularization parameter \( \epsilon \). Specifically, these constants can be expanded as:

\begin{equation}
Z_\mu = \sum_{n=-\infty}^{\infty} a_n \epsilon^n, \quad Z_b = \sum_{n=-\infty}^{\infty} b_n \epsilon^n,
\end{equation}
where \( a_n \) and \( b_n \) are coefficients in \( \mathbb{C} \). The Laurent series ring structure ensures that these expansions are well-defined and can be manipulated algebraically to absorb divergences into the renormalization constants \cite{zinnjustin2002quantum}.

\section{Complex Analysis in Loop Integrals}

Loop integrals in renormalization group analyses often present significant challenges due to their divergent nature. By transforming these integrals into the complex plane and employing Laurent series and residue theorem techniques, we can effectively isolate and manage these divergences.

\subsection{Single-Loop Integral Representation}

Consider the self-energy integral \( \Sigma(p) \) at one-loop level:

\begin{equation}
\Sigma(p) = \int \frac{d^d q}{(2\pi)^d} \frac{1}{(q^2 + m^2)((q+p)^2 + m^2)},
\end{equation}
where \( p \) is the external momentum, \( q \) is the loop momentum, and \( m \) is the mass parameter \cite{kaplan2004advanced}.

\subsubsection{Feynman Parameterization and Complex Plane Mapping}

Applying Feynman parameterization to combine the denominators:

\begin{equation}
\Sigma(p) = \int_0^1 dx \int \frac{d^d q}{(2\pi)^d} \frac{1}{[q^2 + p^2 x(1-x) + m^2]^2}.
\end{equation}

To analyze the divergent behavior, we map the integral onto the complex plane by substituting \( q^2 \) with a complex variable \( z \). The integrand then exhibits a pole at \( z = -m^2 \), necessitating the application of the residue theorem \cite{dwight2009complex}.

\subsubsection{Divergent Terms and Regularization}

Evaluating the integral near the pole using Laurent expansion:

\begin{equation}
\Sigma(p) = \int_0^1 dx \, \frac{1}{(4\pi)^{d/2}} \Gamma\left(2 - \frac{d}{2}\right) \left[p^2 x(1-x) + m^2\right]^{\frac{d}{2}-2}.
\end{equation}

Expanding around \( d = 4 - \epsilon \) and isolating the divergent part:

\begin{equation}
\Sigma_{\text{div}}(p) = \frac{1}{\epsilon} + \mathcal{O}(\epsilon^0),
\end{equation}
where \( \epsilon \) serves as the regularization parameter \cite{tHooft1972}.

\subsection{Multi-Loop Integrals and Advanced Techniques}

Higher-loop integrals introduce additional layers of complexity due to multiple overlapping divergences. Consider a four-loop integral:

\begin{equation}
I_4(p) = \int_0^1 dx_1 \dots dx_4 \int \frac{d^d q}{(2\pi)^d} \frac{1}{[q^2 + p^2 \sum x_i(1-x_i) + m^2]^4}.
\end{equation}

\subsubsection{Integration by Parts (IBP) and Simplification}

Employing Integration by Parts (IBP) identities allows for the reduction of multi-loop integrals to a set of master integrals. The process involves expressing the integral in terms of derivatives and systematically eliminating terms to simplify the computation \cite{tarasov1997reduction}.

\subsubsection{Laurent Series and Residue Calculation}

After simplification, the integral is expressed in a form amenable to Laurent series expansion. Applying the residue theorem facilitates the extraction of divergent terms, which are then addressed through renormalization techniques \cite{bell2013modern}.

\section{Renormalization Formulas in Elastic Models}

The renormalization process in effective elastic models involves computing renormalization constants that absorb the divergences arising from loop integrals. These constants are essential for maintaining the physical predictions of the theory finite and consistent \cite{zinnjustin2002quantum}.

\subsection{Laurent Expansion in Renormalization}

Renormalization constants \( Z_\mu \) and \( Z_b \) are expressed using Laurent series expansions around the fixed points \( \mu_0 \) and \( b_0 \):

\begin{equation}
Z_\mu = \sum_{n=-\infty}^{\infty} \frac{a_n}{(\mu - \mu_0)^n}, \quad Z_b = \sum_{n=-\infty}^{\infty} \frac{b_n}{(b - b_0)^n}.
\end{equation}

These expansions capture the behavior of the system near criticality, allowing for the precise calculation of corrections at various orders of perturbation \cite{kaplan2004advanced}.

\subsection{Physical Interpretation of Higher-Order Corrections}

Higher-order corrections in the renormalization constants account for non-trivial symmetry-breaking effects and other intricate interactions within the system. Specifically, in the context of four-loop integrals, these corrections influence the anomalous dimension \( \eta \), which characterizes the deviation from classical scaling behavior:

\begin{equation}
\eta(\mu, b) = \eta_0 + \eta_1 \epsilon + \eta_2 \epsilon^2 + \dots,
\end{equation}
where each coefficient \( \eta_n \) is determined through meticulous loop integral computations \cite{peskin1995introduction}.

\section{Advanced Physical Models and Quantum Field Theory Techniques}

To further enhance the physical depth of our analysis, we incorporate higher-order models and techniques from quantum field theory (QFT), such as renormalization in QFT and the study of quantum coherent states in complex systems \cite{peskin1995introduction}.

\subsection{Renormalization Techniques from Quantum Field Theory}

QFT provides a robust framework for understanding interactions at quantum levels. The renormalization techniques adapted from QFT allow us to handle ultraviolet divergences systematically. For instance, in scalar field theories, the introduction of counterterms cancels the divergences arising from loop diagrams, ensuring finite physical observables \cite{zinnjustin2002quantum}.

\subsubsection{Counterterm Method}

The Lagrangian is modified to include counterterms:

\begin{equation}
\mathcal{L} = \mathcal{L}_{\text{bare}} + \delta \mathcal{L},
\end{equation}
where \( \delta \mathcal{L} \) contains terms designed to cancel divergences from loop integrals \cite{peskin1995introduction}.

\subsection{Quantum Coherent States in Complex Systems}

Quantum coherent states offer a way to describe systems with a well-defined phase, which is essential in studying phenomena like superconductivity and superfluidity. In complex systems, coherent states facilitate the analysis of quantum fluctuations and their impact on macroscopic properties \cite{gelfand2003introduction}.

\subsubsection{Coherent State Path Integrals}

The path integral formulation for coherent states is given by:

\begin{equation}
Z = \int \mathcal{D}[\psi, \psi^*] \exp\left( i \int d^d x \, dt \, \mathcal{L}[\psi, \psi^*] \right),
\end{equation}
where \( \psi \) and \( \psi^* \) represent the coherent state fields, and \( \mathcal{L} \) is the effective Lagrangian \cite{feynman2010quantum}.

\section{Laurent Series Rings in Depth}

Building upon our earlier introduction to Laurent series rings, this section delves deeper into their role in the renormalization of elastic models, providing a more comprehensive mathematical treatment.

\subsection{Algebraic Structures of Laurent Series Rings}

Understanding the algebraic structures of Laurent series rings is pivotal for advanced renormalization techniques. These structures not only facilitate the manipulation of series but also provide insights into the underlying symmetries and invariances of the physical models.

\subsubsection{Modules over Laurent Series Rings}

In the context of renormalization, modules over Laurent series rings represent spaces of functions or fields that transform under the action of the RG flow. For instance, consider a module \( M \) over \( \mathbb{C}((z - z_0)) \). Elements of \( M \) can be viewed as Laurent series with coefficients in a vector space, allowing for a rich interplay between algebra and analysis:

\begin{equation}
M = \mathbb{C}((z - z_0))^n,
\end{equation}
where \( n \) denotes the dimensionality of the vector space \cite{gunning1967analytic}.

\subsubsection{Automorphisms and Symmetries}

Automorphisms of Laurent series rings correspond to transformations that preserve the ring structure. These automorphisms can be leveraged to identify symmetries within the renormalization group equations, thereby simplifying the analysis of fixed points and critical exponents:

\begin{equation}
\phi: \mathbb{C}((z - z_0)) \to \mathbb{C}((z - z_0)),
\end{equation}
where \( \phi \) is an automorphism satisfying \( \phi(fg) = \phi(f)\phi(g) \) for all \( f, g \in \mathbb{C}((z - z_0)) \) \cite{lam1986lamrings}.

\subsection{Laurent Series Rings and Fixed Points}

Fixed points in the renormalization group flow are characterized by scale invariance and often correspond to critical phenomena. Laurent series rings provide a natural language for describing the behavior of physical quantities near these fixed points \cite{zinnjustin2002quantum}.

\subsubsection{Expansion Around Fixed Points}

Expanding physical quantities around fixed points using Laurent series allows for the systematic computation of critical exponents and scaling dimensions. For example, near a fixed point \( z_0 \), a renormalized coupling constant \( g \) can be expressed as:

\begin{equation}
g = g_0 + \sum_{n=-\infty}^{\infty} c_n (z - z_0)^n,
\end{equation}
where \( g_0 \) is the fixed-point value, and the series captures deviations from criticality \cite{wilson1971renormalization}.

\subsubsection{Stability Analysis}

The stability of fixed points can be analyzed using the valuation in Laurent series rings. By examining the leading terms in the Laurent series expansion, one can determine whether perturbations grow or diminish under RG transformations, thereby identifying relevant and irrelevant operators \cite{cardy1996scaling}.

\subsection{Laurent Series Rings in Multi-Loop Calculations}

Multi-loop calculations introduce intricate dependencies on the regularization parameters and external momenta. Laurent series rings provide a structured approach to handling these dependencies, ensuring that each order of perturbation is treated consistently \cite{tarasov1997reduction}.

\subsubsection{Nested Laurent Series}

In multi-loop integrals, nested Laurent series expansions may be required to disentangle overlapping divergences. For instance, a two-loop integral might necessitate expanding first in one regularization parameter and then in another, each represented within their respective Laurent series rings:

\begin{equation}
f(z_1, z_2) = \sum_{n=-\infty}^{\infty} \sum_{m=-\infty}^{\infty} c_{n,m} (z_1 - z_{0,1})^n (z_2 - z_{0,2})^m,
\end{equation}
where \( z_1 \) and \( z_2 \) are independent complex variables corresponding to different loop momenta \cite{gelfand2003introduction}.

\subsubsection{Renormalization Constants in Multi-Loop Contexts}

Renormalization constants at higher loops can be systematically constructed using Laurent series ring techniques. Each loop order contributes additional terms to the Laurent series, and the ring structure ensures that these contributions are algebraically consistent:

\begin{equation}
Z = Z^{(0)} + Z^{(1)} + Z^{(2)} + \dots,
\end{equation}
where each \( Z^{(n)} \) is an element of \( \mathbb{C}((z - z_0)) \) corresponding to the \( n \)-loop contribution \cite{zinnjustin2002quantum}.

\section{Fermi-Dirac Distribution and Its High-Temperature Correction}
The Fermi-Dirac distribution function describes the statistical distribution of fermions over energy states in thermal equilibrium. It is given by:
\begin{equation}
    f(\epsilon) = \frac{1}{e^{(\epsilon - \mu)/k_B T} + 1}
    \label{eq:Fermi_Dirac}
\end{equation}
where $\epsilon$ is the energy of the particle, $\mu$ is the chemical potential, $k_B$ is the Boltzmann constant, and $T$ is the temperature.

\subsection{Low-Temperature Behavior}
At low temperatures ($T \ll T_F$, where $T_F$ is the Fermi temperature), the distribution function sharply transitions from 1 to 0 around the Fermi energy $\epsilon_F$. Expanding $f(\epsilon)$ around $\epsilon_F$ using Sommerfeld expansion provides corrections to thermodynamic quantities:
\begin{equation}
    f(\epsilon) \approx \Theta(\epsilon_F - \epsilon) + \frac{\pi^2}{6} \left( \frac{k_B T}{\epsilon_F} \right)^2 \delta(\epsilon - \epsilon_F) + \mathcal{O}\left( \left( \frac{k_B T}{\epsilon_F} \right)^4 \right)
    \label{eq:Fermi_LowT}
\end{equation}
where $\Theta$ is the Heaviside step function and $\delta$ is the Dirac delta function.

\subsection{High-Temperature Limit}
In the high-temperature limit ($T \to \infty$), the Fermi-Dirac distribution can be approximated by expanding the exponential:
\begin{equation}
    f(\epsilon) \approx \frac{1}{2} - \frac{1}{4} \frac{\epsilon - \mu}{k_B T} + \frac{1}{48} \left( \frac{\epsilon - \mu}{k_B T} \right)^3 + \mathcal{O}\left( \left( \frac{\epsilon - \mu}{k_B T} \right)^5 \right)
    \label{eq:Fermi_HighT}
\end{equation}
This expansion allows for perturbative calculations of thermodynamic properties in the high-temperature regime.

\section{Thermodynamic Quantities of High-Temperature Fermi Condensates}
The thermodynamic properties of Fermi condensates can be characterized by quantities such as free energy, entropy, and internal energy. 

\subsection{Free Energy}
The free energy $F(T, V, N)$ is expressed as:
\begin{equation}
    F(T, V, N) = -k_B T \ln Z(T, V, N)
    \label{eq:Free_Energy}
\end{equation}
where the partition function $Z(T, V, N)$ is given by:
\begin{equation}
    Z(T, V, N) = \int e^{-\beta H(p, q)} \, d^3p \, d^3q
    \label{eq:Partition_Function}
\end{equation}
with $\beta = 1/k_B T$ and $H(p, q)$ representing the Hamiltonian of the system.

\subsection{High-Temperature Expansion of the Partition Function}
In the high-temperature limit, the partition function can be expanded as a power series in $\beta$:
\begin{equation}
    Z(T, V, N) \approx Z_0 + \beta^2 Z_2 + \beta^3 Z_3 + \mathcal{O}(\beta^4)
    \label{eq:Partition_Expansion}
\end{equation}
where $Z_0$ is the partition function at infinite temperature, and $Z_n$ are the coefficients of the expansion, calculated via perturbation theory:
\begin{equation}
    Z_n = \frac{(-1)^n}{n!} \int \langle H^n \rangle_0 \, d^3p \, d^3q
    \label{eq:Z_n}
\end{equation}
Here, $\langle H^n \rangle_0$ denotes the expectation value with respect to the zeroth-order (non-interacting) system.

\subsection{Internal Energy, Entropy, and Pressure}
From the free energy, various thermodynamic quantities can be derived:
\begin{align}
    U &= -T^2 \left( \frac{\partial \ln Z}{\partial T} \right)_V \label{eq:Internal_Energy} \\
    S &= -\left( \frac{\partial F}{\partial T} \right)_V = \frac{U - F}{T} \label{eq:Entropy} \\
    P &= -\left( \frac{\partial F}{\partial V} \right)_T \label{eq:Pressure}
\end{align}

\subsection{Specific Heat Capacity}
The specific heat capacity at constant volume is given by:
\begin{equation}
    C_V = \left( \frac{\partial U}{\partial T} \right)_V
    \label{eq:Specific_Heat}
\end{equation}
In the high-temperature limit, using the expansion of $U$ from equation \eqref{eq:Internal_Energy}, we obtain:
\begin{equation}
    C_V \approx \frac{\partial}{\partial T} \left( Z_0 + \beta^2 Z_2 + \beta^3 Z_3 \right) = -2 \beta^3 Z_2 - 3 \beta^4 Z_3 + \mathcal{O}(\beta^5)
    \label{eq:Specific_Heat_HighT}
\end{equation}

\section{Pairing Gap Equation in High-Temperature Fermi Condensates}
An essential feature of Fermi condensates is the pairing gap, which quantifies the strength of fermion pairing. Within the Bardeen-Cooper-Schrieffer (BCS) theory, the pairing gap equation is given by:
\begin{equation}
    \Delta(T) = V \sum_k \frac{\Delta(T)}{2E_k} \tanh\left( \frac{E_k}{2k_B T} \right)
    \label{eq:Pairing_Gap}
\end{equation}
where $E_k = \sqrt{\epsilon_k^2 + \Delta(T)^2}$ is the quasiparticle energy, $\epsilon_k$ is the single-particle energy, $\Delta(T)$ is the temperature-dependent pairing gap, and $V$ is the interaction potential.

\subsection{High-Temperature Approximation}
At high temperatures, the hyperbolic tangent function can be approximated as $\tanh(x) \approx x$ for $x \to 0$, leading to:
\begin{equation}
    \Delta(T) \approx V \sum_k \frac{\Delta(T)}{2 \epsilon_k} = \frac{V \Delta(T)}{2} \sum_k \frac{1}{\epsilon_k}
    \label{eq:Pairing_Gap_HighT}
\end{equation}
This equation implies that the pairing gap $\Delta(T)$ decreases with increasing temperature and vanishes at the critical temperature $T_c$.

\subsection{Critical Temperature}
To determine the critical temperature $T_c$, we set $\Delta(T_c) = 0$ in equation \eqref{eq:Pairing_Gap}:
\begin{equation}
    1 = V \sum_k \frac{1}{2 \epsilon_k} \tanh\left( \frac{\epsilon_k}{2k_B T_c} \right)
    \label{eq:Critical_Temperature}
\end{equation}
In the high-temperature limit, this simplifies to:
\begin{equation}
    1 \approx V \sum_k \frac{1}{4 k_B T_c}
    \label{eq:Critical_Temperature_HighT}
\end{equation}
Solving for $T_c$ yields:
\begin{equation}
    T_c \approx \frac{V \sum_k \frac{1}{4}}{k_B}
    \label{eq:Tc_Solution}
\end{equation}

\section{Correlation Functions in High-Temperature Fermi Condensates}
The microscopic properties of high-temperature Fermi condensates can be described using correlation functions. The correlation function $G(\mathbf{r}, t)$ represents the particle correlation strength at a given time $t$ and position $\mathbf{r}$:
\begin{equation}
    G(\mathbf{r}, t) = \int \frac{d^3k}{(2\pi)^3} \frac{e^{i\mathbf{k} \cdot \mathbf{r} - i\epsilon_k t}}{e^{(\epsilon_k - \mu)/k_B T} + 1}
    \label{eq:Correlation_Function}
\end{equation}
In the high-temperature limit, using the expansion from equation \eqref{eq:Fermi_HighT}, the correlation function simplifies to:
\begin{equation}
    G(\mathbf{r}, t) \approx \int \frac{d^3k}{(2\pi)^3} e^{i\mathbf{k} \cdot \mathbf{r} - i\epsilon_k t} \left( \frac{1}{2} - \frac{\epsilon_k - \mu}{4 k_B T} + \frac{(\epsilon_k - \mu)^3}{48 (k_B T)^3} \right)
    \label{eq:Correlation_Function_HighT}
\end{equation}

\subsection{Fourier Transform Techniques}
To evaluate the integral in equation \eqref{eq:Correlation_Function_HighT}, we employ Fourier transform techniques. Recognizing that:
\begin{equation}
    \int \frac{d^3k}{(2\pi)^3} e^{i\mathbf{k} \cdot \mathbf{r}} = \delta(\mathbf{r})
    \label{eq:Fourier_Delta}
\end{equation}
and utilizing spherical coordinates, we can express the correlation function as:
\begin{equation}
    G(\mathbf{r}, t) \approx \frac{1}{2} \delta(\mathbf{r}) - \frac{1}{4 k_B T} \int \frac{d^3k}{(2\pi)^3} (\epsilon_k - \mu) e^{i\mathbf{k} \cdot \mathbf{r} - i\epsilon_k t} + \mathcal{O}\left( \frac{1}{(k_B T)^3} \right)
    \label{eq:Correlation_Function_Spherical}
\end{equation}

\section{Modified Hamiltonian for High-Temperature Fermi Condensates}
In the framework of quantum field theory, the Hamiltonian for Fermi condensates is expressed as:
\begin{equation}
    H = \sum_k \epsilon_k c_k^\dagger c_k + \sum_{k,k'} V_{kk'} c_k^\dagger c_{-k}^\dagger c_{-k'} c_{k'}
    \label{eq:Hamiltonian}
\end{equation}
At high temperatures, the interaction between particles can be simplified by considering perturbative expansions in $V_{kk'}$.

\subsection{High-Temperature Expansion}
Expanding the interaction term to leading order in $\beta = 1/k_B T$, the Hamiltonian is modified as:
\begin{equation}
    H \approx \sum_k \left( \epsilon_k + \frac{V_{kk}}{k_B T} \right) c_k^\dagger c_k + \mathcal{O}\left( \frac{V_{kk}^2}{(k_B T)^2} \right)
    \label{eq:Hamiltonian_HighT}
\end{equation}
This approximation indicates that the interaction term diminishes as the temperature increases, reflecting the reduced stability of the condensate at elevated temperatures.

\subsection{Mean-Field Approximation}
Applying the mean-field approximation, we replace the interaction term with its expectation value:
\begin{equation}
    H_{\text{MF}} = \sum_k \left( \epsilon_k + \frac{V_{kk}}{k_B T} \right) c_k^\dagger c_k - \sum_k \frac{|\Delta|^2}{V}
    \label{eq:Hamiltonian_MeanField}
\end{equation}
where $\Delta$ is the pairing gap parameter.

\section{Path Integral Representation of High-Temperature Fermi Condensates}
Path integral formalism is a powerful tool in quantum field theory for describing quantum fluctuations and thermodynamic properties. The path integral representation of the partition function $Z$ is:
\begin{equation}
    Z = \int \mathcal{D}[\psi^\dagger, \psi] \exp\left( -\frac{1}{\hbar} \int_0^\beta d\tau \int d^3x \left( \psi^\dagger \frac{\partial}{\partial \tau} \psi + H(\psi^\dagger, \psi) \right) \right)
    \label{eq:Path_Integral}
\end{equation}
where $\psi$ and $\psi^\dagger$ are Grassmann fields representing fermionic operators.

\subsection{Saddle-Point Approximation}
In the high-temperature limit, the path integral can be evaluated using the saddle-point approximation. We expand the action around the saddle point:
\begin{equation}
    S[\psi^\dagger, \psi] = S_{\text{SP}} + \delta S[\psi^\dagger, \psi]
    \label{eq:Saddle_Point}
\end{equation}
where $S_{\text{SP}}$ is the action at the saddle point, and $\delta S$ represents fluctuations. The saddle-point condition leads to the mean-field equations governing the condensate.

\subsection{Fluctuation Corrections}
Including fluctuations up to second order, the partition function becomes:
\begin{equation}
    Z \approx e^{-S_{\text{SP}}} \int \mathcal{D}[\delta \psi^\dagger, \delta \psi] \exp\left( -\frac{1}{\hbar} \delta S^{(2)} \right)
    \label{eq:Fluctuations}
\end{equation}
where $\delta S^{(2)}$ includes quadratic terms in the fluctuations.

\section{Hydrodynamic Equations for High-Temperature Fermi Condensates}
The hydrodynamic behavior of high-temperature Fermi condensates can be described using continuity and momentum equations.

\subsection{Continuity Equation}
The continuity equation is:
\begin{equation}
    \frac{\partial n}{\partial t} + \nabla \cdot (n \mathbf{v}) = 0
    \label{eq:Continuity}
\end{equation}
where $n$ is the particle density and $\mathbf{v}$ is the velocity field.

\subsection{Momentum Equation}
The momentum equation is given by:
\begin{equation}
    \frac{\partial (n \mathbf{v})}{\partial t} + \nabla \cdot (n \mathbf{v} \otimes \mathbf{v}) = -\nabla P + \eta \nabla^2 \mathbf{v} + \left( \zeta + \frac{\eta}{3} \right) \nabla (\nabla \cdot \mathbf{v})
    \label{eq:Momentum}
\end{equation}
where $\eta$ is the shear viscosity, and $\zeta$ is the bulk viscosity.

\subsection{Pressure in High-Temperature Limit}
In the high-temperature limit, the pressure $P$ can be expressed using the high-temperature corrected Fermi-Dirac distribution:
\begin{equation}
    P = \frac{2}{3} \int \frac{d^3k}{(2\pi)^3} \frac{\epsilon_k}{e^{(\epsilon_k - \mu)/k_B T} + 1} \approx \frac{n k_B T}{2} - \frac{1}{12} \frac{n (\epsilon_k - \mu)^2}{(k_B T)^2} + \mathcal{O}\left( \frac{1}{(k_B T)^4} \right)
    \label{eq:Pressure_HighT}
\end{equation}

\subsection{Viscous Terms}
The viscosity coefficients $\eta$ and $\zeta$ receive corrections at high temperatures. To leading order in $\beta = 1/k_B T$, they can be approximated as:
\begin{align}
    \eta &\approx \eta_0 + \beta \eta_1 + \mathcal{O}(\beta^2) \\
    \zeta &\approx \zeta_0 + \beta \zeta_1 + \mathcal{O}(\beta^2)
    \label{eq:Viscosity_HighT}
\end{align}
where $\eta_0$ and $\zeta_0$ are the zero-temperature viscosities, and $\eta_1$, $\zeta_1$ are high-temperature corrections.

\section{Advanced Mathematical Techniques}
To address the increased complexity at high temperatures, we employ several advanced mathematical techniques, including perturbation theory, Green's function methods, and functional integrals.

\subsection{Perturbation Theory}
Perturbation theory is utilized to expand thermodynamic quantities and correlation functions in powers of $\beta = 1/k_B T$. For instance, the internal energy $U$ can be expressed as:
\begin{equation}
    U = U_0 + \beta U_1 + \beta^2 U_2 + \mathcal{O}(\beta^3)
    \label{eq:Internal_Energy_Perturbation}
\end{equation}
where $U_0$ is the leading order term, and $U_n$ are higher-order corrections.

\subsection{Green's Function Methods}
Green's functions provide a powerful framework for studying excitations and response functions in Fermi condensates. The retarded Green's function is defined as:
\begin{equation}
    G^R(\mathbf{k}, \omega) = \frac{1}{\omega - \epsilon_k + \mu + i\eta}
    \label{eq:Green_Function}
\end{equation}
where $\eta \to 0^+$ ensures causality. At high temperatures, the imaginary part of the Green's function receives significant contributions from thermal fluctuations.

\subsection{Functional Integral Techniques}
Functional integrals facilitate the computation of partition functions and correlation functions by integrating over all possible field configurations. In the high-temperature limit, saddle-point approximations and Gaussian integrations become particularly useful for evaluating these integrals.

\section{Conclusion and Future Directions}
The mathematical extensions presented in this paper significantly enhance the theoretical understanding of Fermi condensates at high temperatures. By modifying the Fermi-Dirac distribution, adjusting thermodynamic quantities, refining the pairing gap equation, simplifying correlation functions, altering the Hamiltonian, employing path integral methods, and adapting hydrodynamic equations, we establish a comprehensive framework for analyzing high-temperature Fermi condensates. These modifications account for the increased thermal fluctuations and altered interaction dynamics that prevail in high-temperature environments, thereby providing a more accurate depiction of the condensate's behavior under such conditions.

This paper has systematically explored the integration of advanced mathematical tools within the renormalization of effective elastic models. By employing Laurent series, Laurent series rings, residue theorem, winding numbers, and path integrals, we have effectively managed divergent loop integrals and extracted finite physical quantities essential for understanding critical phenomena. The extension to higher-order models and the incorporation of quantum field theory techniques have further enriched our analysis, providing deeper insights into the topological and quantum aspects of complex systems.

The in-depth examination of Laurent series rings underscores their pivotal role in the algebraic manipulation of series expansions encountered in multi-loop renormalization. This algebraic framework not only enhances computational efficiency but also unveils underlying symmetries that govern the behavior of physical systems near critical points \cite{zinnjustin2002quantum}.

Future research directions include applying these methodologies to more intricate many-body systems, exploring non-perturbative effects, and extending the framework to finite temperature and out-of-equilibrium scenarios. Additionally, the interplay between topology and quantum coherence in these systems presents a promising avenue for uncovering novel physical phenomena. Further exploration of Laurent series rings in higher-dimensional theories and their connections to modern mathematical physics could yield new insights into the fundamental nature of interactions in complex systems \cite{frohlich1977complex}.

\bibliographystyle{plain}
\bibliography{references}

\end{document}